\documentclass[12pt]{article}

\RequirePackage[colorlinks,citecolor=blue,urlcolor=blue]{hyperref}
\RequirePackage[OT1]{fontenc}
\RequirePackage{amsthm,amssymb,amsfonts,graphicx,subfigure,caption,bm,fullpage,wrapfig,setspace}

\usepackage{amsmath,amssymb,enumerate}
 \usepackage{float}
\usepackage{algpseudocode,algorithm,algorithmicx}
\usepackage{mathrsfs}
\usepackage{multirow}
\usepackage{forest, adjustbox}
\usepackage{bbm}
\usepackage{lscape}
\usetikzlibrary{calc}
\usepackage{relsize}

\tikzset{fontscale/.style = {font=\relsize{#1}}
    }
\numberwithin{equation}{section}
\theoremstyle{plain}
                          \newtheorem{thm}{Theorem}
                          
\theoremstyle{remark}         
\theoremstyle{definition} \newtheorem{defn}{Definition}
                          \newtheorem{exam}{Example}
                          \newtheorem{assmp}{Assumption}

 \usetikzlibrary{calc}
\usepackage{relsize}
\usepackage{xcolor}

\usepackage{tikz}
\usetikzlibrary{shapes.multipart}

\usepackage{natbib}
\newcommand\E {\mathbbm{E}}

\newcommand\bW{\bm{W}}
\newcommand\bB{\bm{B}}
\newcommand\bThe{\bm{\Theta}}

\newcommand\cC{\mathcal{C}}
\newcommand\cB{\mathcal{B}}
\newcommand\cY{\mathcal{Y}}

\newcommand\bd{\bm{d}}

 \definecolor{airforceblue}{rgb}{0.36, 0.54, 0.66}

\providecommand{\keywords}[1]
{
  \small	
  \textbf{\textit{Keywords: }} #1
}

\title{Trustworthy Online Marketplace Experimentation with Budget-split Design}
\author{Min Liu \\ \small{mliu@linkedin.com}  \and Jialiang Mao \\ \small{jimao@linkedin.com} \and Kang Kang \\ \small{kakang@linkedin.com} }
\date{\textit{LinkedIn Corporation, Sunnyvale, CA} \\ \hspace{1mm} \\ \today}						

\begin{document}
\maketitle


\begin{abstract}
Online experimentation, also known as A/B testing, is the gold standard for measuring product impacts and making business decisions in the tech industry. The validity and utility of experiments, however, hinge on unbiasedness and sufficient power. In two-sided online marketplaces, both requirements are called into question. The Bernoulli randomized experiments are biased because treatment units interfere with control units through market competition and violates the “stable unit treatment value assumption”(SUTVA). The experimental power on at least one side of the market is often insufficient because of disparate sample sizes on the two sides. Despite the important of online marketplaces to the online economy and the crucial role experimentation plays in product improvement, there lacks an effective and practical solution to the bias and low power problems in marketplace experimentation. Our paper fills this gap by proposing an experimental design that is unbiased in any marketplace where buyers have a defined budget, which could be finite or infinite. We then provide generalizable system architecture for deploying this design to online marketplaces. Finally, we confirm our findings with empirical performance from experiments run in two real-world online marketplaces.
\end{abstract}


\section{Introduction}
 
 Online experimentation (also known as A/B testing) has become the gold standard for measuring product impact and making business decisions in the tech industry. Companies such as Google, Facebook, Amazon, Microsoft and LinkedIn have all embraced this paradigm and built in-house platforms to enable large-scale online experimentation \citep{10.1145/1835804.1835810, letham2019, Deng2016, Kohavi2012, 10.1145/2783258.2788602}, commercial platforms have also emerged to provide online experimentation as a service to the broader audience\cite{10.1145/3097983.3097992}. Because of the crucial role online experimentation plays in decision making and the scale at which it impacts the business, it is of utmost importance to ensure the unbiasedness and power of experiments. When an experiment is biased, the impact estimate is systematically incorrect and the quality of business decisions is compromised. This is especially detrimental if the bias is systematic. When the experiments do not have enough power, even large impacts are not statistically significant and are incorrectly neglected as "neutral", and when impacts do get detected, the throughput of experiments, therefore speed of innovation, is limited. These considerations are precisely why the majority of online experimentation literature is devoted to either designing unbiased experiments under an interference structure \citep{10.1145/3097983.3098192, DesignandAnalysisofExperimentsinNetworksReducingBiasfromInterference}, or to increasing the power of an unbiased experiment \citep{10.1145/2433396.2433413}.

The optimal strategy for designing unbiased and powerful experiments is highly dependent on the underlying interference structure. The most studied example of interference in the tech industry is that of the online social network \citep{10.1145/3097983.3098192, DesignandAnalysisofExperimentsinNetworksReducingBiasfromInterference}, where each unit is the same type of entity (for example, a member of a social network). Each unit can connect with other units and interact with her connections.

Another important interference structure in online platforms, yet received little coverage, is that of online marketplaces. Unlike in a social network, the marketplace population consists of at least two types of units, sellers who offer goods, services or attention and buyers who purchase them with some currency. Marketplaces are an important part of the online economy, and usually are the products responsible for revenue generation in a tech companies. Therefore, it is important for both tech companies and the health of the online economy that we are able to experiment in online marketplaces unbiasedly and with high power. Some examples of online marketplaces are ads marketplace, where advertisers purchase the attention of platform members in the form of ads impressions \citep{GoogleAds, FacebookAds}; hiring marketplace, where employers purchase the attention of job seekers in the form of job postings; and education platforms, where buyers purchase courses from educators that offer them.

The interference and network structure in online marketplaces present experimentation challenges to both unbiasedness and power. Bernoulli randomized experiments on either the seller-side or buyer-side violate the ``stable unit treatment value assumption”(SUTVA) \citep{doi:10.1198/016214504000001880, Johari2020ExperimentalDI} and are biased. To illustrate why SUTVA is violated, we use seller-side Bernoulli randomized experiments as an example. When a treatment increases the competitiveness of sellers, unless all buyers still have enough budget reserve to purchase more, the increased attractiveness of treatment sellers translates to relatively less attractiveness of control sellers, and the control sales decreases as treatment sales increases. Such interference in marketplaces violates the SUTVA assumption, which states a unit's potential outcome should depend only on the treatment it receives and not on treatments other units receive. Analogous argument explains why buyer-side Bernoulli randomized experiments also violate SUTVA and are biased. In the marketing sphere, this bias is termed the ``cannibalization bias", to highlight the fact that treatment sales is higher than control sales not because total sales grows, but because treatment sales ``cannibalizes" on control sales. Making a decision based on Bernoulli randomized experiments when cannibalization is present results in systematic overestimation of treatment effects. In any real-world marketplace, there are always limits to buyers' budgets and sellers' inventory, therefore both seller-side and buyer-side Bernoulli randomized experiments are subject to the cannibalization bias. 

Beyond the cannibalization bias, marketplace experimentation is also plagued with insufficient experimental power. This happens because the two sides of the market usually consist of different types of entities, and are drastically disparate in numbers. For example, in an Ads marketplace, where buyers are enterprise advertisers or agencies, and sellers are consumer users, the number of buyers \citep{googleAdvertiser} can be three orders of magnitude smaller than the number of sellers \citep{googleStats}. Similarly, in a physical goods marketplace, where buyers are consumers and sellers are more likely to be small businesses, the number of buyers is 60 times larger than the number of sellers \citep{amazonStats}. This disparity in sample size naturally leads to disparity in experimental power, and the insufficient power will be further exacerbated if the side with smaller sample size also has longer tails in metric distributions. This power disparity, or relative insufficiency on one side of the marketplace is undesirable because the marketplace business is only sustainable if both sides continues to be optimized, and one side of the marketplace having significantly lower experimental power means the speed of innovation and optimization is throttled on this side.

Solutions to marketplace experimentation in existing literature roughly follow two approaches: the analysis approach which first models the interference structure and cannibalization bias,  and then apply model-based adjustments to the biased Bernoulli randomized experiments \citep{Johari2020ExperimentalDI, 10.1145/2600057.2602837}; the design approach, which explore alternative designs that are unbiased under weak and verifiable assumptions on the marketplace \citep{doi:10.1080/01621459.2018.1527225, ebayMarket, DBLP:conf/aistats/BasseSL16}. The analysis approach typically requires first simplifying the marketplace to a parametric model, when the modeling assumptions themselves are often untestable. In fact, if the model could accurately capture the marketplace, experimentation is not necessary and treatment effects can be measured through simulations. Some design approaches \citep{doi:10.1080/01621459.2018.1527225, ebayMarket} are successful in achieving low bias under reasonable assumption, but the low bias comes at the cost of even lower power. Our paper contributes to the marketplace experimentation literature by proposing the budget-split design which overcomes both the bias and power challenges. The unbiasedness of the budget-split design requires one easy-to-validate assumption only, which is the buyers have a continuous budget (the budget can be finite or infinite). It is up to 30 times more powerful than alternative designs when deployed in the real-world marketplaces. The implementation and deployment of this design does not require an overhaul of the existing online marketplace infrastructure or the experimentation platform. Rather, only incremental changes to modules of the infrastructure. We note that our design shares some similarities with the budget allocation design in \cite{DBLP:conf/aistats/BasseSL16}. Although both designs split the ``budgets'' in some ways, they are fundamentally different: in \cite{DBLP:conf/aistats/BasseSL16}, an indirect, artificial “budget” (quota) is given to each buyer while the budget-split design considers the direct, natural budget that the buyers define. We also notice a recent blogpost in \cite{stitchFixBlog} that outlines an inventory-split counterpart of the budget-split design. Our formal analysis of the budget-split design can be directly applied to the inventory-split settings with little modification.

The impact of the budget-split design is significant. It has enabled marketplace experimentation free of cannibalization bias; boosted experimental power to detect treatment effects that previously would have been dominated by noise. As a result, marketplace innovation and optimizations are carried out with much higher confidence and velocity. In this paper, we use the ads marketplace as a running example to present the budget-split design. The methodology, system architecture and empirical results can be readily extend to any marketplace where the buyer has a well-defined budget, such as the hiring marketplace. The rest of the paper is organized as follows. In Section 2, the causal inference problem in marketplaces is formally defined using the potential outcome framework. Section 3 introduces the budget-split design contrasts it with alternative designs such as cluster-based randomization and alternating-day design. Section 4 highlights the key considerations in system architecture and deployment of the budget-split design to online marketplaces. The empirical results on unbiasedness and power of the budget-split design are reported in Section 5. Finally, Section 6 concludes the paper with discussions and future research.


\section{Experimentation in online ads marketplace}
\label{sec:oam}
Experiments in an online ads marketplace (OAM) falls into the family of bipartite experiments where experimentation takes place in a bipartite graph, linking two types of entities -- the buyers and the sellers\citep{pouget2019variance, doudchenko2020causal}.  \textit{Sellers} are consumer members who visit the website, offer their attention and create requests for ads impressions, while \textit{buyers} are advertisers that fulfill these requests with ads impressions. Each member visit can result in multiple ads requests to ``sell'' (i.e., the webpage can have multiple slots to show ads impressions during one visit). Each advertiser can create multiple \textit{campaigns} that serve difference purposes, have different budgets, and target different member segments. Without loss of generality, we set our experimental units on the two sides as members and campaigns. Our framework can be applied to units at other granularity levels though. For example, it is common to set member sessions as the seller side units (finer) and advertisers as the buyer side units (coarser).

Suppose the marketplace contains $N$ members and $M$ campaigns, denoted by $i\in [N]=\{1,2,\ldots, N\}$ and $\cC_j, j\in [M]=\{ 1,2,\ldots, M\}$, respectively. For the $j$-th campaign $\cC_j$, let $B_j \geq 0$ be its budget and the member set $I_j \subset [N]$ be its target population. Let $\bm{B}=\{B_1,B_2,\ldots,B_M \}$ denote the budgets of all campaigns. We note that targeting is at the member level. Thus if a member is targeted by a campaign, all requests created by this member are targeted. In addition, the campaigns can have different strategy and behavior, such as the bidding strategy and the pacing strategy. We do not model these details and instead simply represent these nuisances with parameter $\bThe_j$. With these notations, we write the entire OAM as $\mathcal{M} = \{ (\cC_j, I_j, B_j, \bThe_j)^M_{j=1}\}$. Figure~\ref{fig:illustration} gives an illustration of an OAM. 

There are two types of experiments in an OAM classified by the side of the market where the new feature is to be launched. When this feature acts on the seller-side, for example changing the placement of ads from top of the page to right rail of the page, a \textit{member-level experiment} (ME) or a seller-side experiment needs to be performed to test its effect where the randomization units are members. Similarly, when the feature acts on the buyer-side, such as adjusting the bid price of a campaign, we need to conduct a \textit{campaign-level experiment} (CE) or a buyer-side experiment where the randomization units are campaigns. 

In this paper, we use ME to introduce the ideas. The budget-split design we propose also applies to CE. To simplify the demonstration, we defer the discussion of CE to Section~\ref{sec:bs_campaign_level} and focus on ME for now. In an ME, let $W_i\in\{0,1\}$ be the treatment received by member $i$, $\bW = (W_1,\ldots, W_N)$ the treatment assignment vector of all members. Classical designs, such as the completely randomized design, can be directly applied to the seller side: 
\begin{defn}[Member-level completely randomized design]
A member-level completely randomized design is a probability distribution $\eta$ on $\{0,1\}^N$ such that for some $1\leq N_1\leq N$, $\eta(\bW) = 1$ if $\sum^N_{i=1} W_i = N_1$ and $\eta(\bW) = 0$ otherwise.
\end{defn}
Unfortunately, as we show in Section~\ref{sec:cannibalization_bias} and Section~\ref{sec:6.2}, applying this design in OAM without modification and ignoring the buyer side can severely bias the treatment estimation. Similar bias is entailed when applying classical designs on the buyer side and ignoring the seller side.


 \def\mystrut{\vrule height 0.8cm depth 0.8cm width 0pt} 
 \def\mystrutt{\vrule height 0.4cm depth 0.4cm width 0pt} 
 \def\mystruttt{\vrule height 0.2cm depth 0.2cm width 0pt}

 \begin{figure}[!h]
 \begin{center}
 \resizebox{0.55\textwidth}{!}{
  \begin{tikzpicture}
  \node[rectangle split, rectangle split parts=4,
       draw, minimum width=3cm, font=\large, rectangle split part fill={gray!40,white, white,white},
       rectangle split part align={center}] (t1)
    {             {$\mathbf{Adv_1}$}
     \nodepart{two}
                  \mystrut $campaign_{11}$ 
     \nodepart{three}
                  \mystrutt $campaign_{12}$
     \nodepart{four}
                  \mystrutt $campaign_{13}$
 };
 
  \begin{scope}[xshift=7cm, yshift=0.5cm]
    \node[rectangle split, rectangle split parts=5,
          draw, minimum width=2cm,font=\small, rectangle split part fill={blue!15,white, white, white},
         rectangle split part align={center}] (t2)
     {                {$\mathbf{Member_1}$}
       \nodepart{two}
                      $request_{11}$
       \nodepart{three}
                      $request_{12}$ 
       \nodepart{four}
                      $request_{13}$ 
       \nodepart{five}
                      $request_{14}$ }; 
  \end{scope}

\begin{scope}[xshift=0cm, yshift=-5.5cm]
    \node[rectangle split, rectangle split parts=3,
          draw, minimum width=3cm,font=\large, rectangle split part fill={gray!40,white, white,white},
         rectangle split part align={center}] (t3)
     {                {$\mathbf{Adv_2}$}
       \nodepart{two}
                      \mystrutt $campaign_{21}$
       \nodepart{three}
                      \mystrut $campaign_{22}$ }; 
  \end{scope}

\begin{scope}[xshift=7cm, yshift=-3cm]
    \node[rectangle split, rectangle split parts=3,
          draw, minimum width=2cm,font=\small, rectangle split part fill={blue!15,white, white, white},
         rectangle split part align={center}] (t4)
     {                $\mathbf{Member_2}$
       \nodepart{two}
                      $request_{21}$
       \nodepart{three}
                      $request_{22}$  }; 
  \end{scope}
  
  \begin{scope}[xshift=0cm, yshift=-9cm]
    \node[rectangle split, rectangle split parts=3,
           minimum width=4cm,font=\Large,
         rectangle split part align={center}] (t5)
     {                \textbf{$\cdots$}}; 
  \end{scope}

    \begin{scope}[xshift=7cm, yshift=-6cm]
    \node[rectangle split, rectangle split parts=3, draw,
           minimum width=2cm,font=\small, rectangle split part fill={blue!15,white, white, white},
         rectangle split part align={center}] (t6)
     {                $\mathbf{Member_3}$
     	\nodepart{two}
			$request_{31}$
	\nodepart{three}
			$request_{32}$}; 
  \end{scope}

  \begin{scope}[xshift=7cm, yshift=-9cm]
    \node[rectangle split, rectangle split parts=3, 
           minimum width=2cm,font=\Large,
         rectangle split part align={center}] (t7)
     {                \textbf{$\cdots$}}; 
  \end{scope}
  
  \draw[->] (t1.two east) to [out=0, in=180](t2.text west);
  \draw[->] (t1.two east) to [out=0, in=180](t4.text west);
  \draw[->] (t1.two east) to [out=0, in=180](t6.text west);   
  
  \draw[->] (t1.three east) to [out=0, in=180](t2.text west);
  
  \draw[->] (t1.four east) to [out=0, in=180](t2.text west); 
  \draw[->] (t1.four east) to [out=0, in=180](t4.text west);     
    
  \draw[->] (t3.two east) to [out=0, in=180](t4.text west);
 
  \draw[->] (t3.three east) to [out=0, in=180](t4.text west);
  \draw[->] (t3.three east) to [out=0, in=180](t6.text west);
  \draw[->] (t3.three east) to [out=0, in=180](t7.text west);   
\end{tikzpicture}}
\vspace{-0.3cm}
\end{center}
\caption{An illustration of the online ads marketplace. The size of each campaign block represents its budget.}
\label{fig:illustration}
\end{figure}
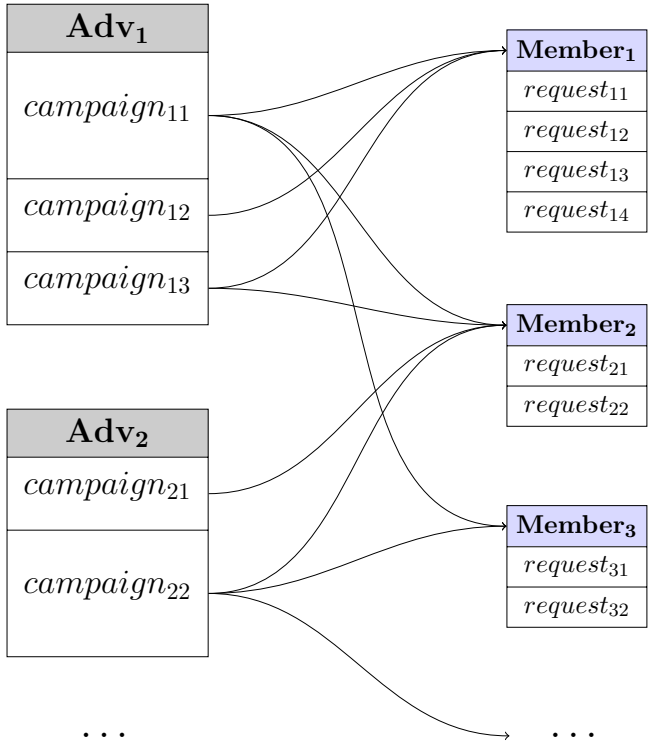


\subsection{Potential outcomes and causal estimands}
\label{sec:po}
We 
adopt the potential outcomes framework \citep{rubin1974estimating} for causal inference. For each $i\in[N]$ and $j\in[M]$, let $Y_{ij}(\bW;\bB)$ be the scalar potential outcome of member $i$ and campaign $\cC_j$ given the treatment assignment vector and the budgets. Note that the potential outcomes are specified for each $(member, campaign)$ pair, which is the finest level in our design. A similar setup is also considered in \cite{basse2016randomization}. This general setup allows outcomes measured at coarser levels to be written as aggregations. For example, for situations in \cite{doudchenko2020causal} and \cite{pouget2019variance} where the potential outcomes are specified in terms of only one side of the market, they can be defined by summing over indexes in $Y_{ij}(\bW;\bB)$ of the other side. Let $\bW_{-i}$ be the $(N-1)$ vector by removing the $i$-th element from $\bW$, we also consider an alternative parametrization of the potential outcomes as suggested by \cite{chin2018central}: $Y_{ij}(\bW;\bB)= Y_{ij}(W_i, \bW_{-i}; \bB)$.

We focus on two metrics of special interests in an OAM: the value delivered to an advertiser and the revenue received by the marketplace operator. Specifically, let $Y_{ij}(\bW;\bB) \geq 0$ be the value delivered to campaign $\cC_j$ (in fact, to the underlying advertiser) by the ads requests created by member $i$. If $i\not\in I_j$, member $i$ is not in the target population of $\cC_j$ and $Y_{ij}(\bW;\bB)=0$ by definition. If $i\in I_j$, $\cC_j$ competes for the ads slots from member $i$'s visits with all other campaigns $\cC_{j^\prime}$ such that $i\in I_{j^\prime}$ by participating in an auction, such as the firce price auction or the generalized second price auction \citep{edelman2007internet}. Note that the potential outcomes depend on the campaign budgets. In an OAM, $\cC_j$'s own budget $B_j$ is used by the system to guide its pacing and bidding strategy \citep{agarwal2014budget}, affecting its own bid underlying each $Y_{ij}$. At the same time, the resulting $Y_{ij}$ also depends on bids from other campaigns, which are themselves determined by their own budgets and actual spending. Unlike in \citep{basse2016randomization}, we directly focus on the outcomes of the auction, ignoring the underlying bids as well as the auction mechanism for generality. 

Aside from the value delivered to the advertisers, another crucial metric is the corresponding revenue received by the marketplace operator, denoted as $Y^*_{ij}(\bW;\bB)$. It is worth noting that $Y^*_{ij}(\bW;\bB)$ may or may not equal $Y_{ij}(\bW;\bB)$. Due to the complex optimization procedure under the hood, $Y_{ij}$ depends on $\bm{B}$ in a ``not exact" manner, meaning that it is possible for $\sum_{i\in I_j}Y_{ij}(\bW;\bB) > B_j$, a phenomenon referred to as ``over-delivery". On the other hand, the platform cannot charge a campaign more than its budget. Therefore, we have for each $j$,
\begin{equation}
\begin{aligned}
0 \leq Y^*_{ij}(\bW;\bB) & \leq Y_{ij}(\bW;\bB) \\
\sum\limits_{i\in I_j} Y^*_{ij}(\bW;\bB) &\leq B_j.
\end{aligned}
\end{equation}

In this work, we conduct inference conditioning on the potential outcomes and take treatment assignment as the only source of randomness. This randomization based inference strategy has the advantage of not entailing any modeling assumptions on $Y_{ij}(\bW;\bB)$. Due to the complex and evolving ecosystem in an OAM, modeling these potential outcomes can be extremely challenging.

We would like to estimate the treatment effect in terms of the total delivered value and the total received revenue conditioning on the budgets:
\begin{equation}
\begin{aligned}
\tau(\bB)  & =\sum\limits_{j\in[M]}  \sum\limits_{i\in I_j} [Y_{ij}(\bm{1};\bB) -Y_{ij}(\bm{0};\bB)] = \sum\limits_{j\in[M]} \tau_j(\bB) \\
\tau^*(\bB) & = \sum\limits_{j\in[M]}  \sum\limits_{i\in I_j} [Y^*_{ij}(\bm{1};\bB) -Y^*_{ij}(\bm{0};\bB)] = \sum\limits_{j\in[M]} \tau^*_j(\bB),
\end{aligned}
\end{equation}
where $\bm{1} = (1,1,\ldots,1)_N$ and $\bm{0} = (0,0,\ldots,0)_N$ denote the all treatment and all control treatment assignment.

The validity of vanilla A/B testing hinges critically on the stable unit treatment value assumption (SUTVA) \citep{imbens2015causal}, which states that (i) there is only a single form of each treatment variant and (ii) a unit's potential outcome is unaffected by the treatment assignment of other units. The no interference assumption, however, is usually violated in marketplace experiments due to the limited campaign budgets. One special scenario where SUTVA at can be justified is in an imaginary world without budget constraints, so that the campaigns are free to bid on any of their targeted members based on their own evaluation of that member's value. This is essentially the SUTVA for bids assumption made by \cite{basse2016randomization}. Formally, we have

\begin{assmp}[No interference under unlimited budget]
If $B_j = \infty$ for each $j\in[M]$, we have $Y_{ij}(\bW; \bm{\infty}) = Y_{ij}(W_i; \bm{\infty})$.
\label{assmp:sutva}
\end{assmp}
In this case, the following simple plug-in estimator is unbiased:
\begin{equation}
\begin{aligned}
\hat\tau_j(\bm{\infty}) &= \sum\limits_{i\in I_j} \frac{W_i\cdot Y_{ij}(\bW; \bm{\infty})}{N_1/N} - \sum\limits_{i\in I_j} \frac{(1-W_i)\cdot Y_{ij}(\bW; \bm{\infty})}{N_0/N},
\end{aligned}
\label{eq:naive}
\end{equation}
where $N_1 = \sum^{N}_{i=1}w_i$, $N_0 = N - N_1$. Moreover, since there is no budget constraint, $Y_{ij}^*(\bW) = Y_{ij}(\bW) = Y_{ij}(W_i)$ and the corresponding plug-in estimator $\hat\tau_j^*(\bm{\infty})$ for  $\tau_j^*(\bm{\infty})$ is also unbiased.


\subsection{Finite budget and the cannibalization bias}
\label{sec:cannibalization_bias}

In reality, it is very unlikely that ads campaigns have unlimited budgets. As a result, Assumption \ref{assmp:sutva} no longer holds and the potential outcomes are affected. With a finite budget, each campaign now needs to take a more ``global" view over all its target members when planning its bid such that the total budget is split to different members wisely. For example, if the treatment has an edge over the control, a wise campaign should take this advantage and allocate more budget to members in the treatment group, resulting in interference between the potential outcomes. Due to this interference, the naive estimator for $\tau_j(\bB)$ and $\tau^*_j(\bB)$ as in (\ref{eq:naive}) is no longer unbiased. To further illustrate this, we consider two examples. The first example illustrate that the naive estimator $\hat\tau_j^*(\bB)$ of $\tau_j^*(\bB)$ is biased in a very common scenario that can happen in a real OAM due to the strict budget constraints. The second example considers a more fundamental case, where even $\hat\tau_j(\bB)$ is biased. From now on, we assume that $0 < B_j < \infty$ for campaign $j\in [M]$.


\begin{exam}[Full budget utilization]
Imagine an OAM that has a single buyer with a \$100 budget, willing to purchase the sellers attention at \$20 per impression; and many sellers each with a different propensity to pay attention to the buyers ads. Naturally, to better serve both the buyer's and sellers' interest, the marketplace improves the algorithm for matching the buyer with sellers who are most interested paying attention to the buyers ads. If the buyer already manages to spend the \$100 budget under sub par seller ranking ($\sum _{i\in[N]}Y_{i1}^*(\bm{0}; 100) = 100$), then improving seller ranking cannot further increase the buyer spending because there simply is no untapped budget (by definition, $\sum _{i\in[N]}Y_{i1}^*(\bm{1}; 100) \leq 100$ and $\tau^*_1(100) \leq 0$). On the other hand, had we run a member-level completely randomized experiment, with sellers in treatment being ranked by a better algorithm, we would have observed a positive treatment effect of the improved algorithm from the plug-in estimator.
\end{exam}

\begin{exam}[Universally beneficial treatment with diminishing returns]
For some campaign $j$ with $I_j = [N]$, suppose that 
{\small\begin{equation}
\begin{aligned}
& \hspace{0.3cm} Y_{ij}(0,|\bW_{-i}| = N - 1;\bB)  \\
&< \cdots \\
& < Y_{ij}(0,|\bW_{-i}| = 1;\bB) \\
&< Y_{ij}(0,\bm{0};\bB)  < Y_{ij}(1,  \bm{1};\bB) \\
& < Y_{ij}(1, |\bW_{-i}| = N-2;\bB) \\
& < \cdots \\
& <Y_{ij}(1, |\bW_{-i}| = 0;\bB),
\end{aligned}
\label{eq:dimreturn}
\end{equation}}
for any $i\in [N]$, where for $0\leq K < N$, $Y_{ij}(W_i, |\bW_{-i}| = K;\bB) = \{Y_{ij}(W_i, \bW_{-i}; \bB):  \sum_{i\in I_j}\bW_{-i} = K \}$. Intuitively, in this scenario, the treatment increases the value of each member. However, members in the treatment group are more valuable when the size of the treatment group is smaller; members in the control group are less valuable when the size of the treatment group is larger. (Consider an analogy to a job market in a county. If a job applicant is the only college graduate, he/she would probably be paid very high; with more and more applicants holding a college degree, the value of such a degree diminishes while those only holding a high school degree can find it harder to even get a job.) In this case, 
\begin{equation}
\begin{aligned}
& \hspace{0.55cm}\E[\hat\tau_j(\bB)] - \tau_j(\bB) \\
& = \left\{ \E\left[ \sum\limits_{i =1}^N \frac{W_i\cdot Y_{ij}(1, |\bW_{-1}| = N_{1} - 1;\bB)}{N_1/N}\right] - \sum\limits_{i=1}^NY_{ij}(1,\bm{1}; \bB)\right\} \\
& + \left\{ \sum\limits_{i=1}^NY_{ij}(0,\bm{0}; \bB) - \E\left[ \sum\limits_{i=1}^N \frac{(1-W_i)\cdot Y_{ij}(0, |\bW_{-1}| = |N_{1}| - 1;\bB)}{N_0/N}\right] \right\}.
\end{aligned}
\label{eq:canni}
\end{equation} 
By (\ref{eq:dimreturn}), the two terms on the RHS are all positive. Thus $\hat\tau_j(\bB)$ has a positive bias.
\end{exam}
Note that in (\ref{eq:canni}), $\sum_{i=1}^NY_{ij}(1,\bm{1}; \bB)$ is overestimated while $\sum_{i=1}^NY_{ij}(0,\bm{0}; \bB)$ is underestimated. We thus refer to the bias of this naive estimator as the \textit{cannibalization bias} as the new version of the product appears to be having an edge over the old one in an experiment, not only because it drives extra values of members in treatment, but also because it ``cannibalises" the members' values in control.

\subsection{Switchback design}
\label{sec:alternating}

One popular strategy to get rid of this cannibalization bias is to treat the entire OAM at different times as the experimental units and adopt the switchback design \citep{bojinov2020design}. Let $t=1,2,\ldots, T$ be $T$ ordered time points. The switchback design assigns random treatment to each $\mathcal{M}_t$. Thus for each fixed $t$, all units (members or campaigns depending on whether the experiment is ME or CE) are assigned to the control or the treatment simultaneously. Since only one treatment is active at any time, there is no cannibalization bias due to cross treatment group intervention.

In practice, applying a switchback design to an OAM can be challenging. Firstly, although the design prevents any cross treatment interventions at any time point, it allows interventions across time points, often referred to as a carryover effect \citep{bojinov2020design}. In an OAM, the level of the carryover effect depends on the features being tested as well as other factors in the ecosystem. In general, it is hard to understand these carryover effects and analysis the switchback experiment properly. Secondly, a switchback experiment can be highly inefficient when it takes sometime for the effect of the feature to become measurable. For example, some features target at improving the performance of a campaign over a week. In this case, the time gap between different assignments has to be at least a week and the overall experimental period can stretch over weeks or even months. Thirdly, there is a conflict between the validity and the efficiency of a switchback experiment. On the one hand, a switchback experiment needs to run long enough to collect enough samples; on the other hand, it implicitly assumes that the OAM does not change much during the experimentation period, which is likely to be unrealistic when this period is long.


\section{Budget-split design}

In this section, we introduce a novel design that removes the cannibalization bias in both the member-level experiment and the campaign-level experiment in an OAM while bypassing the challenges in the switchback design. Our key motivation is that the cannibalization bias comes from interactions between units in different treatment groups. If these interactions are blocked, the cannibalization bias will not exist. Similar to the idea in \cite{ha2020counterfactual}, 
our goal is to simulate the two ``counterfactuals'' of the ads marketplace---one under the control and another under the treatment---with a design, such that a causal relationship can be established by directly comparing these two simulated marketplaces. The success of our strategy depends on two factors: (i) how similar the two simulated marketplaces are before the treatment, and (ii) how similar these simulated marketplaces are to the original marketplace before the treatment. As we shall see, our design automatically ensures (i), that is, the two marketplaces have exactly the same campaigns, very similar members, and the same underlying mechanism (i.e., auction mechanism, bidding strategy, etc.) We state (ii) formally in Assumption~\ref{assmp:average} and validate this assumption with real examples in Section~\ref{sec:6.2}.


Without loss of generality, we assume $I_j=[N]$ for all $\cC_j, j\in[M]$ in this section. Recall that the original OAM is $\mathcal{M} = \{ (\cC_j, I_j, B_j, \bThe_j)^M_{j=1}\}$. We create two independent marketplaces $\mathcal{M}^{(0)}$ and $\mathcal{M}^{(1)}$ out of $\mathcal{M}$ with the following steps:
\begin{enumerate}
    \item perform a random split of the members into two buckets of size $N^{(0)}$ and $N^{(1)} = N - N^{(0)}$. Let $d_i \in \{0, 1\}$ be the bucket indicator of member $i$ and $\bd = (d_1,d_2,\ldots, d_N)$;
    \item split the budget of each campaign $j$ accordingly: \begin{equation}
        \begin{aligned}
        B_j^{(0)} = \frac{N^{(0)}}{N}\cdot B_j &, \quad B_j^{(1)} = \frac{N^{(1)}}{N}\cdot B_j;
        \end{aligned}
        \end{equation}
    \item create a campaign $\cC_{j_0}$ with a budget $B_j^{(0)}$ and a target population $I_{j_0} = \{i: d_i = 0\}$. Let this campaign inherit all features (summarized by $\bThe_j$) from campaign $\cC_j$. Similarly, create a campaign $\cC_{j_1}$;
    \item replace $\mathcal{M}$ with two independent marketplaces: 
    \begin{equation}
      \begin{aligned}
    \mathcal{M}^{(0)} & = \{ (\cC_{j_0}, I_{j_0}, B^{(0)}_j, \bThe_j)^M_{j_0=1}\}\\
     \mathcal{M}^{(1)} & = \{ (\cC_{j_1}, I_{j_1}, B^{(1)}_j, \bThe_j)^M_{j_1=1}\}.
    \end{aligned}
    \end{equation}
\end{enumerate}



Recall that the treatment effect of interests is
\begin{equation}
\begin{aligned}
\tau_j(\bB) = \sum\limits^{N}_{i=1} [Y_{ij}(\bm{1};\bB) - Y_{ij}(\bm{0};\bB)].
\end{aligned}
\end{equation}
Intuitively, under the budget split setup, we use one of the 
$\mathcal{M}^{(l)}$'s to estimate $\sum^{N}_{i=1}Y_{ij}(\bm{1};\bB)$, and use another to estimate $\sum^{N}_{i=1}Y_{ij}(\bm{0};\bB)$. To this end, let $\tilde W_l\in\{0,1\}$ be the treatment indicator of OAM $\mathcal{M}^{l}$, where $\tilde W_1 \sim \text{Bernoulli}(p)$ and $\tilde W_0 = 1-\tilde W_1$. If $\tilde W_l = 1$, we apply the treatment to members in bucket $l$; if $\tilde W_l = 0$, these members will receive the control.

Let $\bB^{(l)} = \{ B_1^{(l)}, B_2^{(l)}, \ldots, B_M^{(l)} \}$ and $\bW^{(l)} = \{W_i: i\in I_{j_l} \}$ for $l=0,1$. Similar to the definition of $Y_{ij}(\bW;\bB)$, we define the potential outcomes associated with campaign $\cC_{j_l}$ as $Y_{ij}^{(l)}(\bW^{(l)}; \bB^{(l)}\mid \bm{d})$ for $l=0, 1$ and $i\in I_{j_l}$. Note that these potential outcomes also depend on $\bd$. (When defining these potential outcomes, we are actually making a no interference assumption, which is formally stated in Assumption~\ref{assmp:bucket}.) Given a split $\bd$, we consider the following estimator:
\begin{equation}
\begin{aligned}
\hat\tau_j^{BS}(\bB) & = \sum\limits_{l=0,1} \tilde{W}_l\sum\limits_{i=1}^N\frac{\mathbbm{1}(d_i = l)Y_{ij}^{(l)}(\bm{1}_{N^{(l)}}; \bB^{(l)} \mid \bd)}{N^{(l)}/N} \\ & - \sum\limits_{l=0,1} (1-\tilde{W}_l) \sum\limits_{i = 1}^N \frac{\mathbbm{1}(d_i = l)Y_{ij}^{(l)}(\bm{0}_{N^{(l)}}; \bB^{(l)} \mid \bd)}{N^{(l)}/N}.
\end{aligned}
\label{eq:bs}
\end{equation}

To study the properties of $\hat\tau_j^{BS}(\bB)$, we first introduce the following definition:
\begin{defn}[proportionally restricted campaign]
For $j\in[M]$, we define a \textit{proportionally restricted version} of $\mathcal{C}_j$ as $\mathcal{C}_j(K)$, for $1\leq K\leq N$, such that $\mathcal{C}_j(K)$ is only allowed to target at $K$ members completely at random with a budget of $\cB_j(K) = KB_j/N$. Moreover, suppose that all campaigns are proportionally restricted in a coupled manner, that is, they are restricted to the same members. Then for each $i\in [N]$, we can define the potential outcomes associated with $\cC(K)$ as $\cY_{ij}(\bW;\cB(K) \mid \bm{d})$, where $\cB(K)=\{\cB_1(K), \ldots, \cB_M(K) \}$, $\bm{d} = (d_1, d_2,\ldots, d_N)$, $d_i$ the indicator of whether member $i$ is in the restricted target population. For convenience, we define $\cY_{ij}(\bW;\cB(K) \mid \bm{d}) = 0$ if $d_i = 0$.
\end{defn}
In practice, this definition is relevant to the throttling procedure that can be applied before the auctions take place \citep{basse2016randomization}. For example, due to resource constraint, the system could not afford to allow each campaign to bid for all its targeting members. We next state two assumptions on the potential outcomes of proportionally restricted campaigns. Assumption~\ref{assmp:bucket} limits the interference to members within the restricted target population. Assumption~\ref{assmp:average} ensures that the restricted campaigns are on average ``similar enough" to the unrestricted ones such that they can be used to perform inference about the original campaigns.
 
\begin{assmp}[Limited interference]
For each $i\in[N]$, $j\in[M]$ and $\bW$, we assume $\cY_{ij}(\bW;\cB(K) \mid \bm{d}) = \cY_{ij}(\bW_{\bm{d} = 1};\cB(K) \mid \bm{d})$, where $\bW_{\bm{d} = 1}$ represents the elements in $\bW$ with $\bm{d} = 1$.  
\label{assmp:bucket}
\end{assmp}

\begin{assmp}[Stable system]
For each $j$ and $1\leq K\leq N$, we assume
\begin{equation}
\begin{aligned}
\frac{1}{N}\sum\limits^{N}_{i=1}Y_{ij}(\bm{1};\bB) & = \E_{\bm{d}}\left[ \frac{1}{K}\sum\limits_{i=1}^N d_i \cY_{ij}(\bm{1}_K;\cB(K) \mid \bm{d}) \right] \\
\frac{1}{N}\sum\limits^{N}_{i=1}Y_{ij}(\bm{0};\bB) & = \E_{\bm{d}}\left[ \frac{1}{K}\sum\limits_{i=1}^N d_i \cY_{ij}(\bm{0}_K;\cB(K) \mid \bm{d}) \right].
\end{aligned}
\end{equation}
\label{assmp:average}
\end{assmp}
Assumption~\ref{assmp:average} characterizes a degree of stability of the system. It requires that a proportionally restricted campaign behaves in a similar manner to its original campaign so that the two created OAMs are similar to the original one. In practice, this assumption is more likely to hold when (i) $K$ is large and (ii) $N$ is large. When $K$ is close to $N$, the $\cC_j(K)$ is essentially $\cC_j$. When $K$ decreases, the ecosystem underlying $\cC_j(K)$ becomes less similar to that of $\cC_j$ and the reliability of this assumption decreases accordingly. In the extreme case when $K$ is close to $1$, $\cC_j(K)$ essentially loses all its flexibility in participating in auctions. As a result, the potential outcomes of $\cC_j(K)$ can be totally distorted from those of $\cC_j$. In reality, there is usually some burn-in threshold $K_0$ such that the system (i.e., the bidding strategies of the campaigns) stabilizes with $K_0$ members. In this case, if $N$ is very large such that $K > K_0$, $\cC_j(K)$ would behave very similarly to $\cC_j$. Fortunately, in OAM, $N$ is usually at least at the scale of millions, in which case $K > K_0$ can be easily satisfied with $K = O(N)$. We note that $K_0$ depends on the level of homogeneity of members. $K_0$ of a market where the members are very similar to each other is generally smaller than $K_0$ in a more heterogeneous market.

\begin{thm}\label{thm1}
Under Assumption~\ref{assmp:bucket} and Assumption~\ref{assmp:average}, if in addition $N^{(0)} = N^{(1)} = N/2$, $p = 0.5$, the budget split estimator in (\ref{eq:bs}) is unbiased: for each $j\in[M]$, 
\begin{equation}
\begin{aligned}
\E_{\tilde W_1, \bd}[\hat\tau_j^{BS}(\bB) ] = \tau_j(\bB).
\end{aligned}
\end{equation}
\end{thm}
The proof is standard and can be found in Appendix B.


\subsection{Campaign-level experiments}
\label{sec:bs_campaign_level}

So far, we have been focusing on ME. Since the budget split framework creates two complete OAM with both the buyer side and the seller side, it can also be used for CE. A CE is involved when the features to be tested act on the buyer-side. Examples of such features include UI changes at the campaign management portal and adjustments to the underlying campaign bidding strategy. 

Similar to the setup in Section~\ref{sec:oam}, let $W^c_j \in \{0, 1\}$ be the treatment received by campaign $\cC_j$, $j\in[M]$ and let $Y^c_{ij}(\bW^c; \bB)$ be the corresponding potential outcomes. Again, we consider the treatment effect in terms of the total delivered value:
\begin{equation}
\begin{aligned}
\tau^c(\bB)  & =\sum\limits_{j\in[M]}  \sum\limits_{i\in I_j} [Y^c_{ij}(\bm{1}^c;\bB) -Y^c_{ij}(\bm{0}^c;\bB)] = \sum\limits_{j\in[M]} \tau^c_j(\bB).
\end{aligned}
\end{equation}
Under the budget split framework, the setup for CE is essentially the same as for ME. In this case, $\tau^c_j(\bB)$ is estimated directly by comparing the two campaigns $\cC_{j_0}$ and $\cC_{j_1}$ with an estimator similar to (\ref{eq:bs}).

Although the same budget-split framework is used for both ME and CE, the resulting experiments have very different nature. In ME, the budget split design essentially performs a completely randomized experiment at the 50-50 split, where the treatment assignments coincide with the budget allocations. Here we adopt the budget split design to remove the cannibalization bias. In CE, the budget split design has a more direct motivation. Namely, we create two (approximately) identical replicates for each experimental units before the experiment and solve the ``fundamental problem of causal inference" \citep{holland1986statistics} by making the two potential outcomes of each unit directly observable. Instead of estimated or imputed, the potential outcomes are \textit{observed} under the budget split design. Therefore, the causal estimands are also observed directly. 
  
For CE, both the switchback design (SB) and the budget split design (BS) aim at creating ``counterfactuals" of any experimental unit (campaign) and making the potential outcomes observable under both treatments. In SB, campaign $\cC_j$ at different time points serve as its ``counterfactuals". In the BS, the ``counterfactuals" are represented by the two half-sized campaigns $\cC_{j_0}$ and $\cC_{j_1}$ that are active simultaneously. When making causal comparisons, SB keeps the size of the campaigns unchanged while BS controls for the time factor.


\section{System Architecture}
\label{sec:sys}
Having presented the budget-split design and highlighted its advantages over member-level experiments, campaign-level experiments and switchback experiments, we now give guidelines for the system architecture changes for deploying this design onto the online ads marketplace. Note that the modules in the ads marketplace are also ubiquitous in other types of marketplaces such as hiring marketplaces and physical goods marketplaces.

To set the context, we introduce the important components in the infrastructure of an online ads marketplace:
\begin{enumerate}

    \item At the core of an online ads marketplace, is an \textbf{ad server} that receives incoming ads requests and responds with ads impressions in real time. Within the ads server, a pacing/bidding module controls the spending speed of each campaign, and eventually when to stop serving impressions. The pacing/bidding module makes serving decisions based on auction signals such as bid price, targeting and auction participation rate.

    \item The \textbf{tracker service} supports the ad server by keeping track of ads events such as impressions, clicks and tallying the campaign-level cumulative spending, along with other performance metrics, such as the click-through rate. Combined with the core ad server, it forms a feedback loop for ads serving, enabling the pacing/bidding module to be built as a feedback controller.

    \item Finally, the marketplace interfaces with advertisers through the \textbf{campaign management system}, where advertiers create and manage their ads campaigns, as well as get billed and reported on campaign performance.
\end{enumerate}

\begin{figure}[h]
  \centering
  \includegraphics[width=0.75\linewidth]{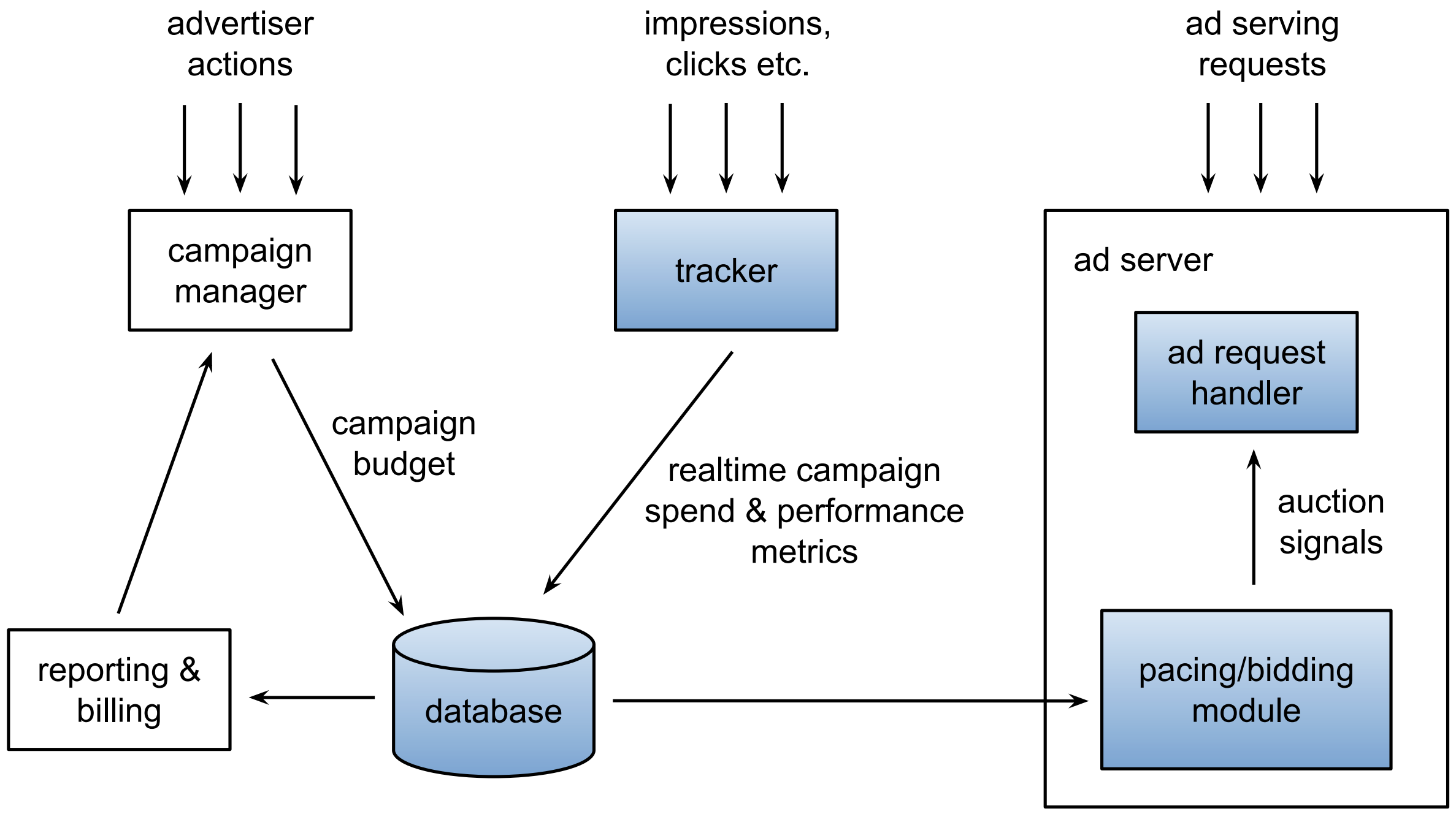}
  \caption{System architecture for the budget-split design.}
  \label{fig:arch}
\end{figure}

To implement the budget-split design, the components marked as blue in figure \ref{fig:arch} are modified, in the following ways:
\begin{enumerate}

    \item In the request receiving and responding tier of the ad server, all requests from a member is randomized into either treatment or control depending on the member ID. This is the member-level randomization as described in Section 3.

    \item In the tracker service, treatment assignment in step 1 together with ads campaign ID are recorded in the tracking information of the request.
    
    \item In the pacing/bidding module of the ad server, all the originally campaign-level controller inputs are now replaced with their campaign-treatment-level counterparts. As an example, budget is now set at campaign-treatment-level as described in Section 3 rather than at campaign-level. Furthermore, feedback signals are provided at per-treatment level. Consequently, the pacing/bidding controller outputs auction signals at the campaign-treatment-level.
    
    Then, the real time auctions, along with any other serving decisions such as whether an ad should stop or resume serving, are carried out using the campaign-treatment-level signals in lieu of campaign-level ones.
\end{enumerate}

With these changes in place, the feedback data and control paths for each treatment become entirely independent of other treatments, for each campaign and across all campaigns. Effectively every treatment now owns a standalone marketplace. This is exactly an implementation of the budget-split design as defined in Section 3.


\section{Empirical Results}
The budget-split experimentation has since been deployed to two online marketplaces. Similar to what was proven in Section 3, it was shown to be much more powerful than alternative designs such as campaign-level experiments or switchback experiments; and not susceptible to cannibalization bias as campaign-level and member-level experiments are.
\subsection{Power Gain}
In the two marketplaces where budget-split was deployed, we compared the power curves of budget-split experiments versus campaign-level and switchback experiments. Switchback experiments were analyzed with paired permutation tests to boost the power; all other designs are analyzed with two-sample t-tests.
\begin{figure}[h]
  \centering
  \includegraphics[width=0.6\linewidth]{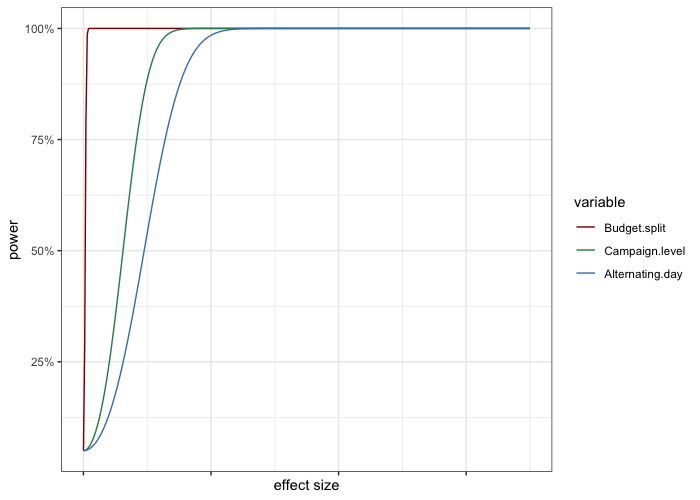}
  \caption{Power curves for marketplace 1}
  \label{fig1}
\end{figure}
\begin{figure}[h]
  \centering
  \includegraphics[width=0.6\linewidth]{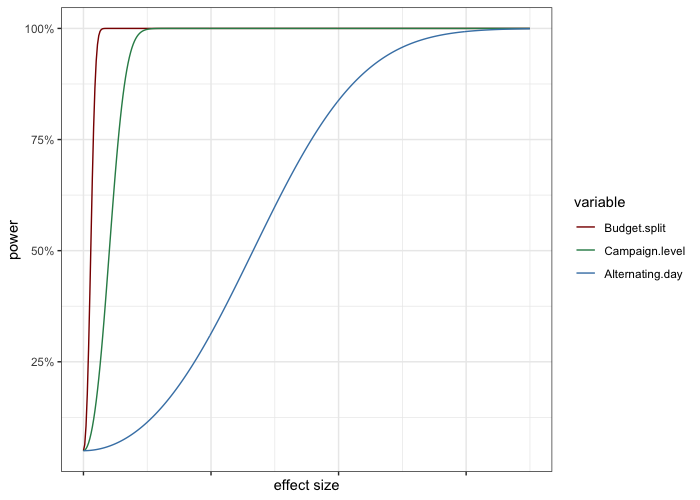}
  \caption{Power curves for marketplace 2}
  \label{fig2}
\end{figure}

As shown in Figure \ref{fig1} and Figure \ref{fig2}, for the effect size that budget-split has 80\% power to detect, campaign-level experiments have power 5.2\% (Marketplace 1) and 12\% (Marketplace 2), while switchback experiments have only 5.1\% (Marketplace 1) and 5.2\% (Marketplace 2) power to detect the effect.

Even though switchback experiments are expected to be unbiased when no carry-over effect exists in the marketplace, they practically have such low power that detecting treatment effects becomes challenging. It is worth noting that 5\% power is actually the lowest power possible for a 5\% level test. Campaign-level experiments are not only biased, but also just barely more powerful than switchback experiments. In practice, budget-split experiments have enabled detection of negative product impacts on the order of millions of dollars (in terms of annual revenue) that previously went undetected in a campaign-level experiment because of insufficient power.

\subsection{Robustness against Cannibalization Bias}\label{sec:6.2}
As discussed in Section 3, the budget-split design effectively removes interference between treatment units and control units through splitting the campaign budgets, and is not susceptible to the cannibalization bias even in place of market competition. In comparison, switchback experiments are unbiased when no carry-over effect is present; cluster-based randomization is unbiased when there is no interference among clusters, which often is not practically feasible\cite{10.1145/3097983.3098192}; member-level and campaign-level experiments are biased because they do not account for the interference from market competition at all. In this section, we compare the impacts measured in budget-split design versus in member-level experiments to showcase the amount of cannibalization bias. It is worth noting that budget-split design was not compared against campaign-level experiments, switchback experiments or cluster-based experiments because these designs are too insufficiently powered for comparison.
\begin{figure}[h]
  \centering
  \includegraphics[width=0.6\linewidth]{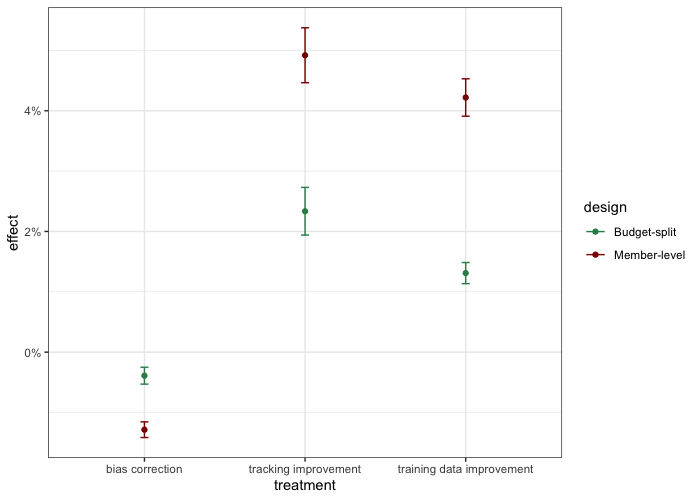}
  \caption{Measurement of Cannibalization Bias}
  \label{fig3}
\end{figure}

The presence of cannibalization bias in marketplace experiment when interference are not accounted for is confirmed in Figure \ref{fig3}. The exact amount of bias depends on the nature of the treatment, but not accounting for interference typically results in one to two times overestimation of the treatment effect. We have not yet encountered a treatment where cannibalization bias changes the sign of the treatment effect, but this is not impossible if, for example, the treatment positively impacts budget-constraint campaigns and negatively impacts non budget-constraint campaigns.


\section{Discussions}
In this paper, we presented the budget-split design as the solution to cannibalization bias and insufficient power in marketplace experimentation. Rather than relying on modeling assumptions that are often impossible to validate, the budget-split design only requires one side of the marketplace having a defined and splittable budget, and no additional assumptions. We formulated the marketplace experimentation problem under the potential outcome framework, defined the estimand, derived an unbiased estimator for the estimand under budget-split design. We showed the bias and variance reduction nature of budget-split design relative to alternative designs with empirical results.

This work, however, is only the start of experimentation in online marketplaces. Future work is need to accurately compute the variance estimators under the budget-split design; extend the budget-split design to marketplaces where the budget is discrete; and increase throughput of budget-split experiments by enabling multiple simultaneous orthogonal budget-split experiments in the marketplace.


\section*{Acknowledgements}
We would like to thank Weitao Duan, Shan Ba, Xingyao Ye and Giorgio Martini for their careful review and feedback; Wei Wei, Qing Duan, Di Luo for implementing the budget-split platform; Shahriar Shariat, Vangelis Dimopoulos, Giorgio Martini and Junyu Yang for contributing to experimental and metric designs and analyses; and everyone who utilized and helped improve the platform. We would also like to thank Ya Xu, Parvez Ahammad, Le Li and Jerry Shen for their continued support. The authors declare no conflict of interest.


\bibliography{budget_split_arxiv}


\clearpage

\section*{Supplementary Materials}

\renewcommand{\thesubsection}{\Alph{subsection}}

\renewcommand{\thefigure}{S\arabic{figure}}
\renewcommand{\theequation}{S\arabic{equation}}

\subsection{Illustration of the cannibalization bias}
\begin{figure}[h]
  \centering
  \includegraphics[width=0.6\linewidth]{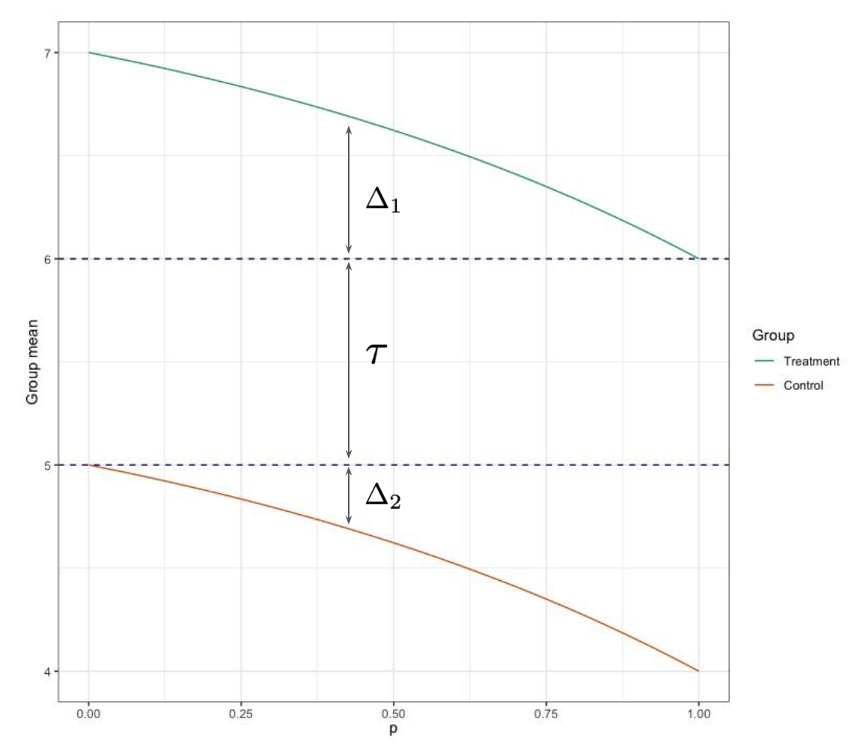}
  \caption{Example of the cannibalization bias}
  \label{fig:cbias2}
\end{figure}

As a further illustration to the cannibalization bias, we consider a special case of Example 2 in Section~\ref{sec:cannibalization_bias}. For simplicity, we take a super-population perspective and assume the potential outcomes are generated by the following model:
\begin{equation}
\begin{aligned}
Y_{i1} (\bW; B_1) = \mu + \tau W_i -\gamma\cdot \frac{e^{\frac{\sum^{N}_{j=1}W_j}{N} - 1} - 1/e}{1-1/e} + \epsilon_i,
\end{aligned}
\end{equation}
where $\mu = 5$, $\tau = 2$, $\gamma = 1$ and $\mathbbm{E}[\epsilon_i] = 0$. In a completely randomized experiment with $N_i = pN$, Figure~\ref{fig:cbias2} shows the expected group means for the two treatment groups. In this example, we have a constant cannibalization bias $\gamma$, which equals $\Delta_1 + \Delta_2$ in Figure~\ref{fig:cbias2}. Note that when $\tau = 0$, the real treatment effect is zero and we have an illustration of Example 1 in Section~\ref{sec:cannibalization_bias}.


\section{Proof of Theorem \ref{thm1}}
\proof 

Under the budget-split design, the two sub-campaigns created out of campaign $j$ are essentially two proportionally restricted versions of $j$. Specifically, we have  
\begin{equation}
\begin{aligned}
Y_{ij}^{(l)}(\bm{1}_{N^{(l)}}; \bB^{(l)} \mid \bd) & = \begin{cases}
& \cY_{ij}(\bm{1}_{N^{(1)}};\cB(N^{(1)}) \mid  \bm{d}),\quad l = 1, d_i = 1 \\
& \cY_{ij}(\bm{1}_{N^{(0)}};\cB(N^{(0)}) \mid  1-\bm{d}) , \quad l = 0, d_i = 0\\
\end{cases} \\
Y_{ij}^{(l)}(\bm{0}_{N^{(l)}}; \bB^{(l)} \mid \bd) & = \begin{cases}
& \cY_{ij}(\bm{0}_{N^{(1)}};\cB(N^{(1)}) \mid  \bm{d}),\quad l = 1, d_i = 1 \\
& \cY_{ij}(\bm{0}_{N^{(0)}};\cB(N^{(0)}) \mid  1-\bm{d}) , \quad l = 0, d_i = 0 \\
\end{cases}.
\end{aligned}
\end{equation}
Therefore, 
\begin{equation}
\begin{aligned}
& \hspace{0.55cm}\E_{\tilde W_1, \bd}\left[ \sum\limits_{l=0,1} \tilde{W}_l\sum\limits_{i=1}^N\frac{\mathbbm{1}(d_i = l)Y_{ij}^{(l)}(\bm{1}_{N^{(l)}}; \bB^{(l)} \mid \bd)}{N^{(l)}/N} \right] \\
& = \E_{\bd}\left[ \frac{1}{2}\sum\limits_{l=0,1}  \sum\limits_{i=1}^N\frac{\mathbbm{1}(d_i = l)Y_{ij}^{(l)}(\bm{1}_{N^{(l)}}; \bB^{(l)} \mid \bd)}{1/2}  \right]  \\
& = \E_{\bd}\Bigg[ \sum\limits_{i=1}^N \mathbbm{1}(d_i = 0)Y_{ij}^{(0)}(\bm{1}_{N^{(0)}}; \bB^{(0)} \mid \bd)  \\
& \hspace{0.9cm}+  \sum\limits_{i=1}^N  \mathbbm{1}(d_i = 1)Y_{ij}^{(1)}(\bm{1}_{N^{(1)}}; \bB^{(1)} \mid \bd)  \Bigg] \\
& = \E_{1-\bd}\Bigg[  \sum\limits_{i=1}^N \mathbbm{1}(1- d_i = 1) \cY_{ij}(\bm{1}_{N^{(0)}};\cB(N^{(0)}) \mid  1-\bm{d})   \Bigg ] \\
& \hspace{0.3cm} +   \E_{\bd}\left[  \sum\limits_{i=1}^N \mathbbm{1}(d_i = 1) \cY_{ij}(\bm{1}_{N^{(1)}};\cB(N^{(1)}) \mid \bm{d})   \right ]  \\
& = \frac{1}{2} \sum\limits^{N}_{i=1}Y_{ij}(\bm{1};\bB)  +\frac{1}{2} \sum\limits^{N}_{i=1}Y_{ij}(\bm{1};\bB) \\
& =  \sum\limits^{N}_{i=1}Y_{ij}(\bm{1};\bB) .
\end{aligned}
\end{equation} 
Similarly, 
\begin{equation}
\begin{aligned}
& \hspace{0.55cm}\E_{\tilde W_1, \bd}\left[ \sum\limits_{l=0,1} (1-\tilde{W}_l) \sum\limits_{i=1}^N\frac{\mathbbm{1}(d_i = l)Y_{ij}^{(l)}(\bm{0}_{N^{(l)}}; \bB^{(l)} \mid \bd)}{N^{(l)}/N} \right] \\
& =  \sum\limits^{N}_{i=1}Y_{ij}(\bm{0};\bB).
\end{aligned}
\end{equation} 
\hfill$\square$

\end{document}